# Diverse and widespread contamination evident in the unmapped depths of high throughput sequencing data

Richard W. Lusk

Department of Ecology and Evolutionary Biology, University of Michigan, Ann Arbor, Michigan, United States of America

Email: rlusk@umich.edu

**Background**

Trace quantities of contaminating DNA are widespread in the laboratory environment, but their presence has received little attention in the context of high throughput sequencing. This issue is highlighted by recent works that have rested controversial claims upon sequencing data that appear to support the presence of unexpected exogenous species.

**Results**

I used reads that preferentially aligned to alternate genomes to infer the distribution of potential contaminant species in a set of independent sequencing experiments. I confirmed that dilute samples are more exposed to contaminating DNA, and, focusing on four single-cell sequencing experiments, found that these contaminants appear to originate from a wide diversity of clades. Although negative control libraries prepared from 'blank' samples recovered the highest-frequency contaminants, low-frequency contaminants, which appeared to make heterogeneous contributions to samples prepared in parallel within a single experiment, were not well controlled for. I used these results to show that, despite heavy replication and plausible controls, contamination can explain all of the observations used to support a recent claim that complete genes pass from food to human blood.

**Conclusions**

Contamination must be considered a potential source of signals of exogenous species in sequencing data, even if these signals are replicated in independent experiments, vary across conditions, or indicate a species which seems *a priori* unlikely to contaminate. Negative control libraries processed in parallel are essential to control for contaminant DNAs, but their limited ability to recover low-frequency contaminants must be recognized.

**Introduction**

While contamination by foreign DNA is a concern for many experiments, it requires particular attention for those that rely on sensitive methods to describe samples that are themselves dilute or degraded. Poor quality samples can be outcompeted by contaminant DNAs over the course of an experiment, and methods to characterize and eliminate contaminants have been rigorously evaluated in fields where such samples are common. These studies have found that free DNA at trace but detectable levels is widespread, often being found in "clean" new PCR tubes [1], dNTPs [2], and a variety of other sources [3-8]. Even extreme precautions, such as UV-treatment of reagents, the use of positive pressure laboratory ventilation systems, etc., appear sufficient only to reduce, rather than eliminate, the abundance of contaminant DNAs [9-11].

These results were established using PCR, but they have received surprisingly little attention in the context of high throughput sequencing, which shares with PCR the ability to potentially detect single DNA molecules. Moreover, sequencing samples are processed through a longer and more complex experimental pipeline, which typically includes a PCR amplification step, increasing their exposure to contaminants. The power of sequencing to recover, but not necessarily to identify, contaminant DNAs should grow as read depths increase and as library preparation methods require progressively smaller amounts of input material. Nevertheless, the presentation of contaminants in sequencing data is poorly understood. Although sequences from humans and a handful of other species have been found to contaminate sequencing samples and databases [12-16], the roster of potential contaminant species is unknown, as is their distribution within and between experiments, making it difficult to infer whether or not any given read originated from the intended sample.

To help create a framework for this inference, I sought to describe the diversity of contaminant reads in four independent sequencing experiments and how these reads are replicated across samples and negative controls. I used this information to evaluate whether contaminants may have evaded the heavy replication and plausible controls described in a recent paper to provide an alternative explanation for its claim that complete genes pass from food to human blood.

**Results and discussion**

*Contaminant read count is inversely related to sample concentration*

High throughput sequencing quantifies the relative, not absolute, quantities of different DNAs in a library. If we assume that the amount of contaminant DNA contributed to each of a set of identically-prepared libraries is constant, we should expect that the number of reads matching contaminant genomes will increase as the concentration of the sample decreases. To illustrate this relationship, I reanalyzed data from a study that sequenced different dilutions of a single sample of *E. coli* DNA [17], examining how changes in the concentration of *E. coli* DNA affected the number of reads matching sequences in the human genome, a common laboratory contaminant [12,14].

The authors of this study sought to demonstrate the efficacy of a novel low-concentration library preparation protocol. To this end, they used it to create libraries from 1ng, 100pg, and 10pg of a shared *E. coli* DNA sample; they also used 1µg of this DNA to prepare a sequencing library using the standard Illumina protocol. As expected, there is an inverse relationship between the concentration of the sample and the frequency of contaminant reads (Figure 1A). For the four libraries prepared from the highest concentration of DNA (1ng), I found only three reads in total that mapped to the human but not to the *E. coli* genome. However, the frequency of these reads increased with decreasing sample concentration, to approximately 175 and 2,500ppm in the 100 and 10pg samples, respectively ($z=-108.4$, $p<2e-16$, see Methods). The protocol used to prepare the libraries can influence the amount of contamination, however: the protocol described above used a single tube to fragment input DNA, ligate adapters, and amplify fragments, potentially reducing the sample's exposure to contaminants. I found an intermediate frequency of contaminating reads in the library prepared from the standard Illumina protocol, despite its higher concentration of sample DNA (Figure 1B).

*Unmapped reads from independent experiments match a wide range of species*

Although contamination by DNA from humans [12,14] and a handful of bacterial species [13,15,16] has been described in sequencing data, there has been to my knowledge no systematic effort to describe the diversity of contaminants evident in sequencing data, making it difficult to gauge the likelihood that any given species inferred to be present from sequencing data originated from contamination. To address this issue, I performed a metagenomic analysis of potential contaminants in a set of four independent sequencing experiments [18-21].

In order to recover the greatest possible yield of contaminant DNAs, I chose experiments that worked with low quantities of input material. All of these experiments prepared libraries from individual cells, but they otherwise had different goals and used different species and experimental techniques (Table 1). For each of these experiments I used permissive settings to screen out reads that could potentially map to the appropriate reference genome, and I then used BLAST to search for perfect matches to the remaining reads in the NR database across the entire or, depending on the dataset, close to the entire length of each read (see Methods). As highly conserved regions are less informative for taxonomic classification, I further screened out reads with reported BLAST hits to more than one broad taxonomic category.

In all of these experiments I found reads matching sequences belonging to a wide diversity of groups (Figure 2), and the observed abundance of each group was broadly correlated across experiments. These groups include some which might seem unlikely contributors to contamination, such as Streptophyta, a phylum encompassing the land plants with and green algaes.

There are several potential sources for reads such as these that preferentially map to alternative genomes in sequencing experiments. In some cases, they may truly be evidence for an alternate organism in the sample, e.g. the serendipitous sequencing of new genomes of the bacterial endosymbiont *Wolbachia* in several *Drosophila* genome sequencing projects [22]. This seems unlikely for the reads described in Figure 2, because all of these experiments sequenced individual cells. Many of these reads map to organisms that seem likely to contaminate samples, such as *Propionibacterium acnes*, a pervasive skin bacterium. Other reads appear to map to alternate genomes due to sequencing errors. While most reads containing errors will presumably not perfectly match any species' genome, the large number of reads analyzed here ensures that many will coincidentally match genomes of species only related to those actually present in the sample. For instance, many reads in each experiment matched the chimpanzee genome, even though the sample organism was either mouse or human. Whether a given match to an exogenous species originated from a true contaminant or from a sequencing error can be difficult to distinguish. Indeed, some of the matches to the chimpanzee genome, particularly those from the mouse experiments, may be due to errors made during sequencing of contaminant human DNA. To reflect this ambiguity, I will refer to reads that match exogenous genomes as "contaminant matching reads."

*'Blank' negative controls effectively describe the distribution of contaminants per sample*

The "RNAseq" and "Strandseq" (Table 1) experiments prepared negative control libraries from "blank" buffer samples into which no sample cell had been deposited. However, only a small number of such libraries were prepared, and each produced a relatively small number of reads, having in total 0.56% and 0.21% of the corresponding total number of reads in the positive samples from the RNAseq and Strandseq studies, respectively. It is possible that this low yield compromised their ability to represent the spectrum of contaminants found in the positive samples.

To address this possibility, I compared the frequency distribution of genera represented by contaminant matching reads in the positive and blank samples, discarding reads that promiscuously matched more than one genus. I found that the frequency distribution of recovered genera correlated well between positive and blank samples from the same experiment, suggesting that, despite their low yield, blank samples can serve as effective negative controls (Figure 3). Furthermore, the correlation is poorer between mismatched positive and blank samples taken from different experiments, consistent with the spectrum of contaminants being experiment-specific. For instance, reads matching the skin bacterium *Propionibacterium acnes* are more than 100-fold more frequent in the genomic DNA sample than they are in the RNA-seq dataset, perhaps reflecting the stricter guidelines for handling RNA.

Despite the strong correlation between the positive and blank libraries, the low yield of the blank libraries limited their ability to control for low frequency genera in the positive libraries, many of which were not represented in the blanks. To lower this detection threshold, it may be sufficient to increase the number of blank samples or to use a distinguishable carrier DNA to increase yield (e.g. [23]), although the potential for introducing additional contaminants in the carrier DNA must be considered.

*Libraries prepared from a single tissue appear contaminated by different species*

If a reagent or piece of equipment is heavily contaminated by a given species, then all exposed samples should share evidence of being contaminated by that species. In this scenario, one way to control for contamination would be to examine how exogenous species are distributed, considering as contaminants species that are recovered from most or all samples. However, while this assumption is reasonable if the contaminant concentration is high, we should expect a weaker correlation between exogenous species if the contaminant concentration is low. Indeed, at the lower limit, where each contaminant species contributes one DNA molecule, samples will necessarily be contaminated by different species. To examine whether these low-frequency contaminants are detectable and to what extent they differ between samples, I reanalyzed data from the "Tumor" experiment in which independent sequencing samples were prepared from the same tissue rather than from independent individuals. In this case, all differences between samples outside changes to the target genome can be attributed to contamination.

In this experiment, Navin et al [18] used FACS to isolate 100 individual nuclei from a section of a single tumor, and, processing them similarly through 96-well plates, used whole genome amplification to accumulate sufficient DNA for library preparation and sequencing. For each of these libraries, I first screened out loose matches to the human genome before searching for perfect matches to a database of chloroplast genomes. I found reads matching chloroplast DNAs in all samples, and different samples had different rosters of inferred contaminants (Figure 4; Additional File 1, Table S1), suggesting that contaminants can contribute unevenly to samples within a single experiment. Many of the species recovered are edible, such as tomato and lettuce, although an appreciable fraction matched to algal or otherwise inedible species (e.g. oak).

*Contaminants provide an alternative explanation for observations of plant DNAs in human blood*

Highlighting the potential impact of contaminants in high throughput sequencing data is a controversial paper that claimed that complete genes pass from food to human blood [24]. The authors of this study based their claim on their observations of chloroplast sequences from edible plants in sequencing data from samples of blood plasma. They were able to replicate this observation in independent datasets, and, contrary to what might be expected from a single contamination event, found that the plant species recovered differed from sample to sample within a single experiment. Furthermore, while these reads could not be detected in a negative control sample of fetal blood, which circulates independently, they found many in the corresponding maternal plasma sample, which is presumably more exposed to the digestive system.

On the basis of their observation's heavy replication, heterogeneity among samples, absence from a plausible negative control, and, finally, their assumption that plant DNA would be unlikely to infiltrate stringent laboratory practices, the authors dismissed the possibility of contamination. However, as I have shown above, all of these assumptions about the nature of contamination are incorrect. DNA from plants is a common contaminant, and it should be expected to appear to replicate across independent experiments (Figure 2). Furthermore, the particular species inferred can vary considerably from sample to sample within an experiment (Figure 4). Finally, although the authors did find hundreds of reads matching plants in a maternal plasma sample, and none in a matched sample of whole fetal blood, the concentration of DNA in whole blood is tens-of-thousands-fold higher than that in blood plasma (approximately 60 µg/ml vs. 21ng/ml [25,26]). Given the inverse relationship between sample DNA concentration and contaminant frequency (Figure 1), we should expect plasma samples to exhibit more evidence of contamination than full blood samples. Indeed, the corresponding maternal whole blood sample contained only one plant-matching read, a number statistically indistinguishable from zero, suggesting that using fetal whole blood as a negative control in this case was inappropriate.

**Conclusions**

Contaminants can in some cases effectively mimic behavior intuitively expected from true signals, including replication across independent experiments and variation between samples, and can include species that are not typically considered potential contaminants. This calls into question several controversial works that have rested their claims on observations of rare matches to exogenous species in sequencing data [24,27-29], and suggests that blank negative controls, prepared in parallel with experimental samples, are essential. Nevertheless, the rarest contaminants, being difficult to recover in these controls, may be intrinsically difficult to control for.

**Materials and methods**

*Inference of human contaminants in* E. coli *sequencing data*

I downloaded data from Parkinson et al [17] from the European Nucleotide Archive. Run accession numbers are listed in Additional File 2, Table S2. Splitting apart the paired-end reads, I used bowtie version 1.0.0 [30] to attempt to map each end to the *E. coli* K12 reference genome, allowing up to three mismatches in the seed region ('-n 3') but otherwise using default parameters. Unmapped reads were then aligned to the human hg18 reference genome [31] with no mismatches permitted in the seed ('-n 0'). Reads were considered a 'hit' if both ends from each pair mapped to within one kilobase of each other. The fraction of human reads was considered the total number of paired end reads meeting these criteria over the total number of pair-end reads. A generalized linear model was used to describe the relationship between the log of the sample DNA concentration and the frequency, modeled with the binomial distribution, of human-matching reads within the total library. This analysis was performed using the GLM2 package in R [32,33].

*Inference of contaminant reads and species composition in four sequencing experiments*

I downloaded data from the short read archive corresponding to study accession numbers SRP002535 (referred to in the main text as the "Tumor" dataset [18]), SRP006834 ("RNAseq" [20]), SRP017186 ("Sperm" [21]), and SRP014866 ("Strandseq" [19]). The specific SRA accessions for each dataset are listed in Additional File 3, Table S3. Due to differences in scope and sequencing technology, the numbers of reads sequenced by these studies differed over orders of magnitude, and so samples from the RNAseq and Strandseq datasets were pooled to provide a more comparable number of total reads, whereas samples having close to the median number of reads in the Sperm and Tumor datasets were analyzed individually. Given the unusual library preparation of the RNAseq dataset [20], in which each read contains a barcode and an uncertain number of guanines, I determined the information content of each position in the library (Additional File 5, Figure S1) and, based upon these data, removed the first fifteen base pairs of each read. The Strandseq and Sperm studies used paired-end libraries; I split mate pairs and processed each end independently.

For each dataset, I used bowtie as described above to screen out reads that aligned to the appropriate mouse or human reference genome (hg18 [31] for Sperm and Tumor; mm9 [34] for RNAseq and Strandseq), allowing three mismatches in the seed region. Using the Amazon EC2 image ami-ef42d586, which contains snapshots of the NCBI nucleotide databases from 02/04/2013, I used BLAST to search for hits to each unaligned read to the nr database, using flags –task megablast, -outfmt 6, and –num_alignments 5 (500 for the negative control analysis in Figure 4). Each experiment having different read lengths and qualities, I considered hits of 100% identity over 48, 75, 40, and 101 base pairs for the Tumor, Strandseq, RNAseq, and Sperm datasets, respectively, to be a perfect match. Due to its large size, the number of BLAST runs for the Strandseq dataset was capped at 7.5 million.

For each BLAST match, I used the GI sequence identification number to locate the corresponding taxID from the NCBI taxonomy database, discarding reads that did not cross-reference. For each taxID, I descended the tree to the fourth-lowest node and assigned the BLAST hit to that taxonomic category. Reads were discarded if they had BLAST hits to more than one of these categories, if they matched the sequences from the species of the reference genome, or if they only matched sequences in the database that were taxonomically unclassifiable (e.g. taxID 155900, "uncultured organism"). Reads matching these criteria were also used for Figure 3, where the taxonomy tree was descended to the "genus" level and reads that matched to more than one genus were discarded.

*Diversity of chloroplast DNAs in individual libraries*

Using the sequencing data described in Figure S3 and the methods described above, I screened out reads in each of these 100 samples that matched the human hg18 genome. I then used bowtie to attempt to match each unmapped read to a database of chloroplast genomes (Additional Table 4, Table S4), allowing no mismatches in the seed region ('-n 0'). These settings were specified to closely follow those used by the Spisak et al study [24].

**Competing interests**

The author declares that he has no competing interests.

**Description of additional data files**

Additional data file 1 is a table listing, as described in Figure 4, the total number of hits to each chloroplast genome in each library from the Tumor dataset. Additional data file 2 is a table listing the European Nucleotide Archive accession numbers associated with each sample analyzed in Figure 1. Additional data file 3 is a table listing the Sequence Read Archive accession numbers associated samples analyzed in Figures 3 and 4. Additional data file 4 is a table listing the chloroplast genomes which made up the chloroplast genome database used in Figure 4. Additional data file 5 is a PDF figure depicting the information criterion used to trim reads in the RNAseq datatset.

**Acknowledgements**

I would like to thank Patricia Wittkopp, Brian Metzger, Kraig Stevenson, Joseph Coolon, and Fabien Duveau for critical comments on the manuscript. I would also like to thank Vincent Denef, in addition to those listed above and other members of the Wittkopp laboratory, for valuable discussions. This work was supported by NIH postdoctoral fellowship F32-GM100685.

**Illustrations and figures**

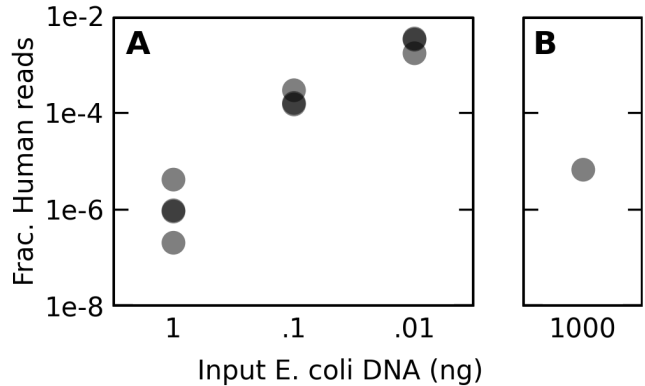

**Figure 1. Reads matching the human genome are more prevalent in libraries prepared from dilute samples.** (a) The fraction of paired-end reads which preferentially map to the contaminant human genome instead of the *E. coli* K-12 genome, measured against the total number of reads in the library, is plotted against the amount of *E. coli* K-12 DNA used per tagmentation procedure as described by Parkinson et al [17]. Shading is used to highlight closely overlapping points (n=4, 3, and 3 for the 1ng, 100pg, and 10pg libraries, respectively). Libraries listed at each concentration were not identically prepared, each using a different restriction enzyme or set of restriction enzymes at an intermediate step in the protocol (Additional File 2, Table S2), but the number and composition of enzymes used did not appreciably change the number of contaminant reads recovered. (b) The same fraction is plotted for a library prepared in the same experiment using a standard Illumina library preparation protocol. Despite a higher concentration of input DNA, an intermediate number of contaminant reads was detected.

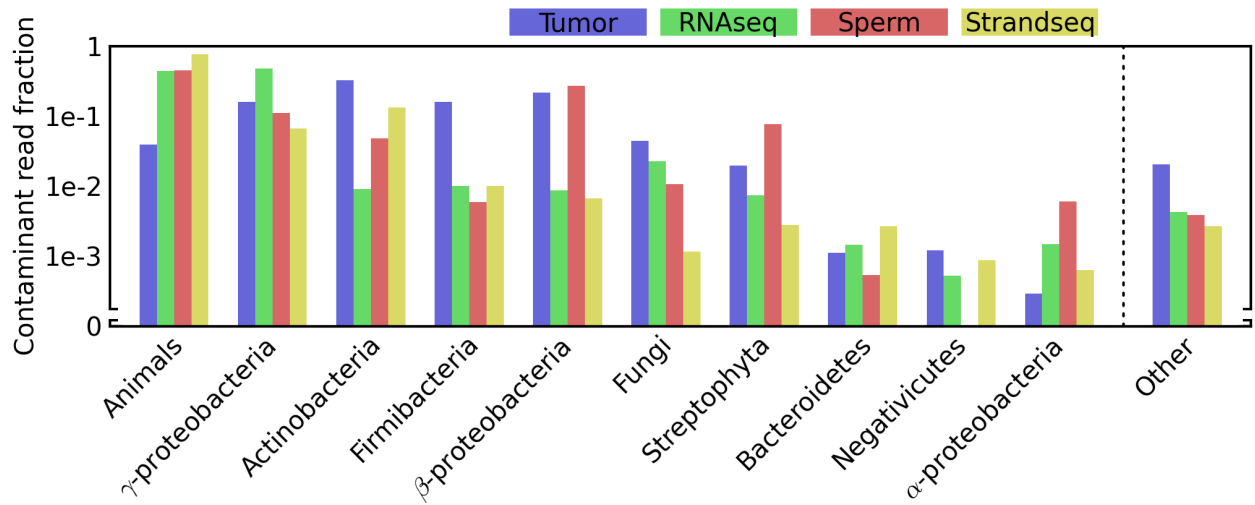

**Figure 2. Reads that do not map to the reference genome match a diverse array of clades.** For each experiment ("Tumor" [18], "RNAseq" [20], "Sperm" [21], and "Strandseq" [19]), all reads were individually mapped to the appropriate reference genome using permissive parameters before being used to query the NR database using BLAST. BLAST hits were considered "perfect" if they matched with 100% identity over a dataset-specific length threshold (see Methods). A read was assigned to one of the depicted phylogenetic categories if it did not map or have a perfect BLAST hit to the reference genome, had a perfect BLAST hit to a species in that category, and had no BLAST hits to species outside that category. For each category and experiment, the fraction of reads meeting this criteria against the total number of reads in the experiment is depicted.

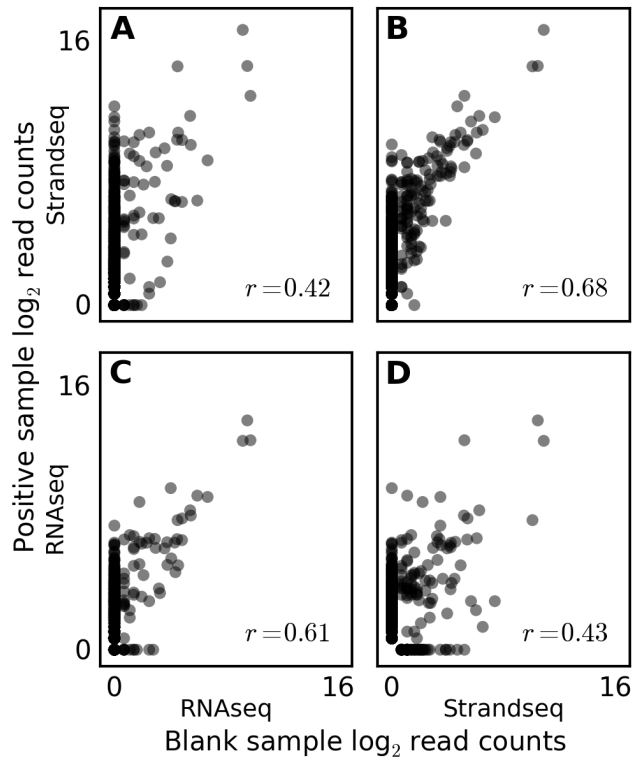

**Figure 3. Experiment-specific correlation between distributions of genera recovered from 'blank' and positive samples.** The "RNAseq" [20] and "Strandseq" [19] experiments sequenced libraries prepared from blank samples into which no cells had been introduced. Reads from these blank samples and from all other samples were separately pooled, screened against the mouse reference genome, and queried against the BLAST NR database. Reads were screened using the same criteria as described for Figure 2, but adjusted to the genus taxonomic level. The number of reads matching each genus in each dataset was counted, incremented by one, and log transformed. Values for the pooled positive samples (StrandSeq and RNAseq in rows A-B and C-D, respectively) are plotted with their Pearson correlation against values for the pooled negative samples (Strandseq and RNAseq in columns A-C and B-D, respectively). Matched positive and negative samples

in (B) and (C) exhibit more correlated read counts than do mismatched positive and negative samples in (A) and (D).

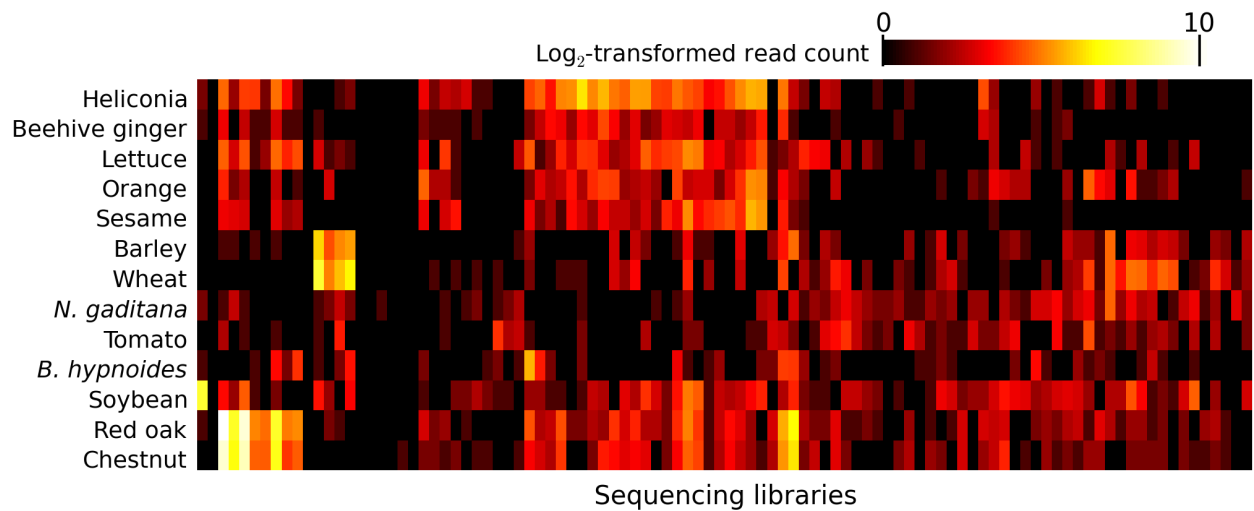

**Figure 4. Heterogeneous species appear to contaminant samples from the same tissue and experiment.** The "Tumor" [18] experiment dissociated 100 individual cells from a sample of a single tumor and sequenced libraries from each. Following the analysis pipeline of a study that claimed to find different plant species in different blood plasma samples from a single experiment, I used bowtie to screen each read in each library against the human reference genome before using it to query a database of chloroplast genomes. The number of such hits to each genome is depicted here, each count incremented by one and log-transformed. Only chloroplast genomes with at least 200 hits are shown. Rows and columns were clustered using a neighbor-joining algorithm.

**Tables and captions**

**Table 1. Description of the sequencing experiments used in this study.**

| Experiment | Reference | Organism | RNA/DNA | Cell isolation | Pooled | Number of reads |
|---|---|---|---|---|---|---|
| RNAseq | [20] | Mouse | RNA | Cell picker | Yes | 93,471,748 |
| Tumor | [18] | Human | DNA | FACS | No | 8,568,573 |
| Strandseq | [19] | Mouse | DNA | FACS | Yes | 100,026,526 |
| Sperm | [21] | Human | DNA | Pipetting | No | 17,066,385 |